\documentclass[11pt]{amsart}
\usepackage{setspace,url,hyperref}
\usepackage{amssymb,stmaryrd}
\usepackage{proof}
\usepackage[usenames,dvipsnames,svgnames,table]{xcolor}

\theoremstyle{definition}

\theoremstyle{remark}

\numberwithin{equation}{section}

\def\InputModeColorName{MidnightBlue}
\def\OutputModeColorName{Maroon}
\newcommand\InputMode[1]{{\color{\InputModeColorName}{#1}}}
\newcommand\OutputMode[1]{{\color{\OutputModeColorName}{#1}}}

\newcommand\HypJ[2]{#1\ (#2)}
\newcommand\GenJ[2]{\vert_{#1}\; #2}
\newcommand\JJ{\mathcal{J}}
\newcommand\IsVer[1]{\InputMode{#1}\ \mathit{ver}}
\newcommand\IsTrue[1]{\InputMode{#1}\ \mathit{true}}
\newcommand\IsProp[1]{\InputMode{#1}\ \mathit{prop}}
\newcommand\IsSet[1]{\InputMode{#1}\ \mathit{set}}
\newcommand\EqSet[2]{\InputMode{#1}=\InputMode{#2}\ \mathit{set}}
\newcommand\Member[2]{\InputMode{#1}\in \InputMode{#2}}
\newcommand\EqMember[3]{\InputMode{#1}=\InputMode{#2}\in \InputMode{#3}}
\newcommand\Seq[2]{\InputMode{#1}\vdash #2}

\newcommand\Eval[2]{\InputMode{#1}\Rightarrow \OutputMode{#2}}
\newcommand\IsDefined[1]{\InputMode{#1}\ \mathit{defined}}

\newcommand\SEM[1]{\llbracket#1 \rrbracket}
\newcommand\VAL[1]{\mathcal{V}\SEM{#1}}
\newcommand\EXP[1]{\mathcal{E}\SEM{#1}}

\newcommand\Nat{\mathbb{N}}
\newcommand\True{\top}
\newcommand\False{\bot}
\newcommand\Imp[2]{#1\supset #2}
\newcommand\Conj[2]{#1\land #2}
\newcommand\Disj[2]{#1\lor #2}
\newcommand\Forall[3]{(\forall #2 \in #1)#3}
\newcommand\Exists[3]{(\exists #2 \in #1)#3}
\newcommand\It{\star}
\newcommand\Lam[2]{(\lambda #1) #2}
\newcommand\Pair[2]{\langle #1, #2 \rangle}
\newcommand\Inl[1]{\mathsf{inl}(#1)}
\newcommand\Inr[1]{\mathsf{inr}(#1)}

\newcommand\BISH{\textbf{BISH}}

\begin{document}

\title{Remark on the hypothetical judgment}

\author{Jonathan Sterling}
\address{}

\onehalfspacing

\begin{abstract}

What is the proper explanation of intuitionistic hypothetical judgment, and
thence propositional implication? The answer is unclear from the writings of
Brouwer and Heyting, who in their lifetimes propounded multiple (sometimes
conflicting) explanations of the hypothetical judgment. To my mind, the
determination of an acceptable explanation must take into account its adequacy
for the expression of the bar theorem and, more generally, the development of
an open-ended framework for transcendental arguments in mathematics.

\end{abstract}

\maketitle

\section{Judgments and Propositions}

The distinction between the propositions and the judgments (assertions) is an
old one, but prior to Martin-L\"of, the significance of assertions was limited
to the affirmation of the truth of propositions. Following Martin-L\"of
\cite{siena.lectures}, forms of judgment other than $\IsTrue{P}$ are recognized,
including $\IsProp{P}$.

What is the difference between a judgment (assertion) on the one hand, and a
proposition on the other hand? A judgment is an act or an experience, whereas a
proposition is a mathematical object which may be experienced in different ways.
For instance, the assertion of the truth of a proposition (i.e.\ $\IsTrue{P}$)
consists in the fulfillment of the intention expressed by the proposition, while
the recognition of an object as a proposition (i.e.\ $\IsProp{P}$) is the act of
understanding this intention.

In addition to the categorical judgments above, higher-order forms of judgment
are also explained, including the hypothetical judgment and the general
judgment. Now, the primitive hypothetical judgment
$\HypJ{\JJ_2}{\JJ_1}$,\footnote{Hypothetical judgment is to be distinguished
from the sequent judgments $\Seq{\Gamma}{\cdots}$, which are not even
higher-order judgments at all.} was explained by Martin-L\"of in terms of
hypothetical \emph{proof} or \emph{demonstration}, which he defined as follows:

\begin{quote}
  The notion of hypothetical proof [demonstration], in turn, which is a
primitive notion, is explained by saying that it is a proof [demonstration]
which, when supplemented by proofs [demonstrations] of the hypotheses, or
antecedents, becomes a proof [demonstration] of the thesis, or
consequent.\footnote{
I prefer the term \emph{demonstration} to the more ambiguous \emph{proof}, since
the former is clearly an act, whereas the latter may be read as either an act or
as a mathematical object.
}
\cite{siena.lectures}
\end{quote}

In 1956, Heyting propounded his version of what has come to be known as the
\emph{Brouwer-Heyting-Kolmogorov} interpretation of intuitionistic logic, by
explaining the assertion conditions of the propositions. Note that where Heyting
says ``assert a proposition'', in light of Martin-L\"of's clarification, we must
read ``assert \emph{the truth of} a proposition''. Heyting's explanation
of the assertion conditions for the truth of implication were as follows:
\begin{quote}
  The implication $\Imp{\mathfrak{p}}{\mathfrak{q}}$ may be asserted if and only
if we possess a construction $\mathfrak{r}$, which, joined to any construction
proving $\mathfrak{p}$ (supposing the latter be effected), would automatically
effect a construction of $\mathfrak{q}$. \cite{HeytingA:int}
\end{quote}

Now, Martin-L\"of would probably consider the parenthetical ``supposing the
latter be effected'' to be superfluous, since to assume a judgment is the same
as to assume that you know it \cite{siena.lectures}.
So, Heyting's definition might be rewritten today as:

\begin{quote}
$\IsTrue{\Imp{P}{Q}}$ may be asserted if and only if we
possess a construction $\mathfrak{r}$, which, joined to any demonstration of $\IsTrue{P}$,
would automatically effect a demonstration of $\IsTrue{Q}$.
\end{quote}

Martin-L\"of explained the truth of an implication by appealing to the
hypothetical judgment, so we should be able to factor Heyting's explanation
through it in a similar way:

\begin{quote}
  $\IsTrue{\Imp{P}{Q}}$ may be asserted if and only if we may assert $\HypJ{\IsTrue{Q}}{\IsTrue{P}}$.
\end{quote}

In fact, if we make this transformation, we shall have arrived at something very
similar to Martin-L\"of's definition of propositional implication. This
inference is, at least, valid with respect to Martin-L\"of's definition, but it
is merely an extensional specification for the meaning of the judgment: it
expresses the material equivalence of the assertions $\IsTrue{\Imp{P}{Q}}$ and
$\HypJ{\IsTrue{Q}}{\IsTrue{P}}$, but it does not contain an actual
\emph{explanation} of $\IsTrue{\Imp{P}{Q}}$, which would need to be in the form ``To
know $\IsTrue{\Imp{P}{Q}}$ is to know...''.

It is tempting to rewrite the definition in the following way:
\begin{quote}
  (*) To know $\IsTrue{\Imp{P}{Q}}$ is to know $\HypJ{\IsTrue{Q}}{\IsTrue{P}}$.
\end{quote}

However, as a definition, this is impredicative. Following Dummett
\cite{Dummett:Elements} and Martin-L\"of, we must start from a distinction
between the direct experience of truth (which we shall call \emph{verification})
and the indirect experience (which we shall just call \emph{truth}). Then, the
intention of a proposition is its verification, and the truth of a proposition
is the experience of a means of fulfillment for that intention:

\begin{quote}
  To know $\IsVer{\Imp{P}{Q}}$ is to know $\HypJ{\IsTrue{Q}}{\IsTrue{P}}$.
\end{quote}

Then, to know $\IsTrue{\Imp{P}{Q}}$ is to have a means of verifying $\Imp{P}{Q}$, that
is, to have a plan to experience $\IsVer{\Imp{P}{Q}}$.

\section{The Proof Interpretation}

When we rewrote Heyting's explanation of the assertion conditions for the truth
of an implication to appeal to hypothetical judgment, we implicitly assumed that
the instantiation of Martin-L\"of's hypothetical judgment would preserve the
meaning of the original statement.

This, however, may be too much to ask, since in modern proof-theoretic accounts
of meaning, the hypothetical judgment as explained by Martin-L\"of must be
understood in a very strong sense, where its proof shall be an object with a
``hole'' in it, which could be plugged with a proof for the antecedent to yield
a proof for the consequent: that is, it is not enough that one should have a way
of transforming the proof of the antecedent into a proof of the consequent, but
one must have a uniform way to do so.

Anticipating the ``proof interpretation'' of intuitionistic logic, Brouwer also
had come to a conclusion similar to this, if not quite equivalent. For Brouwer,
a hypothetical assertion $\HypJ{\JJ_2}{\JJ_1}$ was essentially an assertion of
$\JJ_2$ which proceeds by embedding an actual construction of $\JJ_1$ into a
matrix for a construction of $\JJ_2$. Now, this is not quite the same since it
requires that $\JJ_1$ be proved, so it corresponds more closely with the notion
of \emph{cut} than it does with hypothetical judgment---but he does seem to
agree as far as the uniformity of the partial proof of the consequent is
concerned.

In case it is not entirely evident, let us reason through what would happen to a
\emph{proof theory} if proofs of hypothetical judgments could be non-uniform.
For one, it would cease to be a proof theory, since two crucial properties would
fail:

\begin{enumerate}
  \item Proofs are finitary objects.
  \item It is effectively determinable in finite time whether an object is
    a proof of a judgment.
\end{enumerate}

If a proof of $\HypJ{\JJ_2}{\JJ_1}$ were construed as a means of converting
proofs of $\JJ_1$ into proofs of $\JJ_2$ (as opposed to the proof-with-a-hole
interpretation), proofs would certainly cease to be finitary objects: for
instance, a proof of $\GenJ{n}{\HypJ{\JJ(n)}{\Member{n}{\Nat}}}$ would be
infinitely large, in that it would contain one branch for each natural number.
Then, the failure of the second property (decidability of wellformedness) is
immediate.

Proof-theoretic meaning, however, is hopeless anyway as an interpretation of
intuitionistic logic if one has any intention to justify the creating subject
\cite{mva:personal}, or even the bar theorem; indeed, the proof-theoretic
reading of hypothetical judgment specifically rules out the kind of non-uniform
evidence that is essential in the demonstration of the bar theorem.

\section{The Proper Interpretation of Hypothetical Judgment}

If we return to Heyting's original definition of the assertion
$\IsTrue{\Imp{P}{Q}}$, it is clear that he would have to accept any mathematical
means of transforming the construction of $\IsTrue{P}$ into a construction of
$\IsTrue{Q}$, since the construction $\mathfrak{r}$ in his definition is not
constrained by any uniformity condition: it is merely any effective operation
which, when adjoined with a construction of the premise, effects a construction
of the conclusion; and moreover, $\mathfrak{r}$ may proceed by appealing to any
of the transcendental observations which are possible by virtue of the truth of
$P$ having been experienced by the subject.

And what of Martin-L\"of? In his type theory, only the introduction rules for
the types are given, and the elimination rules are ``theorems'' (or admissible
rules) which are evident only in the non-uniform sense; that is, the
justifications of the elimination rules proceed by introspection on the possible
ways in which their premises could have been experienced. Therefore, if
Martin-L\"of's uniform hypothetical judgment is to be accepted, it is necessary
that the statement of the elimination rules be effected using something that
permits a non-uniform demonstration.

Sundholm and Van Atten \cite{sundholm-mva}, for instance, distinguish between a
hypothetical proof of $\HypJ{\JJ_2}{\JJ_1}$ and an inference
$\infer{\JJ_2}{\JJ_1}$ which expresses the closure of mathematics under a rule,
and say that it is the latter which establishes the bar theorem, and not
the former.

Martin-L\"of on the other hand says specifically that an inference is to be read
as a proof of a hypothetical judgment, and so no real progress is made:
\begin{quote}
  The difference between an inference and a logical consequence, or hypothetical
  judgement, is that an inference is a proof of a logical consequence. Thus an
  inference is the same as a hypothetical proof. \cite{siena.lectures}
\end{quote}

In light of Martin-L\"of's explanations of the elimination rules which follow
this statement, we cannot accept his claim that inference is the same as
hypothetical proof (as he has defined it), because under that definition even
the following is not a valid inference:
\[
  \infer{
    \IsTrue{P}
  }{
    \IsTrue{P\land Q}
  }
\]

Why not? This purports to be a proof of the judgment
$\HypJ{\IsTrue{P}}{\IsTrue{P\land Q}}$, which is not evident under the uniform
explanation of hypothetical judgment. It could be made evident if the definition
of a proposition consisted in the declaration of both its introduction rules and
its elimination rules, but in fact, only the introduction rules are given, and
the elimination rules are simply codifications of common patterns of reasoning
from premise to conclusion.

Indeed, Martin-L\"of justifies the elimination rules using the non-uniform
(material, rather than logical) consequence. The above rule, for instance, is
justified as follows:

\begin{proof}
If you know $\IsTrue{P\land Q}$, then you must have a means of verifying $P\land
Q$, whence you must know both $\IsTrue{P}$ and $\IsTrue{Q}$; the conclusion is
now immediate.
\end{proof}

If we are to take the explanation of the elimination rules seriously, then, we
must read Martin-L\"of as having already at his disposal a kind of hypothetical
judgment whose evidence consists in any effective means at all of transforming
the demonstration of the premise into a demonstration of the conclusion. So, the
use of the uniform hypothetical judgment elsewhere has not relieved us from the
need to explain the inference from premise to conclusion in an elimination rule.

Contra Martin-L\"of's explanation in the Siena lectures, the interpretation of
hypothetical judgment as material consequence is crucial for the semantics of
his type theory, as noted by Dybjer \cite{dybjer:testing}.

\subsection{What is the difference between an inference rule and a hypothetical judgment?}

Is a rule of inference really the same as a consequence or hypothetical
judgment, as Martin-L\"of claimed? One way to elucidate the differences
is to consider them in the context of a Beth or Kripke semantics.

The validity of an inference rule $\infer{\JJ_2}{\JJ_1}$ at a world lies in an
effective transformation of experiences of $\JJ_1$ at that world to experiences
of $\JJ_2$ at that world; on the other hand, to experience a hypothetical
judgment $\HypJ{\JJ_2}{\JJ_1}$ at a world is to have a means to transform
experiences of $\JJ_1$ at any future world into experiences of $\JJ_2$.

Construed in this way, judgments can be explained by specifying when/where they
are forced; it is a reasonable requirement that if we shall consider $\JJ$ to
be a judgment, then for any worlds $u\preceq v$, from $u\Vdash\JJ$ we may
conclude $v\Vdash\JJ$ (this is called \emph{monotonicity}). The hypothetical
judgment, at least as explained above, preserves the monotonicity of knowledge
by definition, whereas rules of inference may not in general satisfy this property.

On the contrary, an ``admissible rule'' $\infer{\JJ_2}{\JJ_1}$ is sensitive to
changes in the state of knowledge, and may cease to be valid if a previously
unknown way to experience $\JJ_1$ is found. Such rules may only be construed as
hypothetical judgments if the acts specified by the meanings of their premises
are sufficiently circumscribed so as to satisfy the monotonicity requirement.

Martin-L\"of's identification of the rules of logic with hypothetical judgments, then,
only obtains because the concepts of \emph{verification} of a proposition, and
(secondarily) \emph{truth} of a proposition are fixed in advance for all time
by means of \emph{canonical forms} and \emph{computation} respectively.

\section{Realizability and Type Theory}

Now that we have settled upon an explanation for hypothetical judgment, let us
return to the notions of judgment and proposition, and their respective concepts
of ``construction''. Like ``proof'', the term ``construction'' is also ambiguous
in that it may refer to an act of constructing, and it may also denote a
concrete mathematical object.

A construction for a judgment is simply the act of coming to know it: this is
what Martin-L\"of calls a \emph{demonstration}, and if it is to be thought of as
an object, it is at least a tensed, ephemeral one. On the other hand, a
construction for a proposition is a mathematical object, not an experience: it
is the object that the subject constructs during the verification of a
proposition. This latter sort of construction is called a \emph{witness}, or,
following realizability, a \emph{realizer}.

\subsection{Realizability Models as Unary Logical Relations}

In fact, we can replace the abstract/synthetic explanations of the propositions
in terms of their verification acts with new explanations in terms of
verification objects (i.e.\ their canonical witnesses); then, the verification
act consists in constructing a verification object.

A proposition is verified just in case there exists a verification object, but
it is important to understand that this is not to say that a (possibly unknown)
verification object may exist outside the subject's experience (construction) of
it. Rather, this is a trivial equivalence, since to say that an object exists is
the same as to say that the subject has constructed it.\footnote{The idea that
verification or proof objects exist separately from our experience of them is
part of the realist ontology which is now espoused by Martin-L\"of
\cite{prawitz:2012}, contrary to his position at the time of the Siena lectures;
this view of course cannot be accepted by Brouwerians, who profess a thoroughly
idealist ontology \cite{sundholm-mva, sundholm:2014}.}

A realizability model in this simple sense amounts to interpreting the
propositions into unary logical relations. To define a proposition, then, is to
define the unary relation $\VAL{P}$ (which is the species of verification
objects of the proposition $P$); then, a separate logical relation $\EXP{P}$ is
defined uniformly over all propositions $P$ by appealing to the computation
$\Eval{M}{N}$ of witnesses to canonical form.

\begin{gather*}
  \infer={
    \IsSet{A}
  }{
    \Eval{A}{A'} &
    \IsDefined{\VAL{A'}}
  } \qquad
  \infer={
    \Member{M}{A}
  }{
    \Eval{A}{A'} &
    \EXP{A'}(\InputMode{M})
  }
\end{gather*}
\begin{align*}
  \VAL{\True} &\equiv \{\InputMode{\It}\}\\
  \VAL{\False} &\equiv \{\}\\
  \VAL{\Imp{P}{Q}} &\equiv
    \{ \InputMode{\Lam{x}{E}}
    \mid \GenJ{x}{\HypJ{\Member{E}{Q}}{\Member{x}{P}}}
    \}\\
  \VAL{\Conj{P}{Q}} &\equiv
    \{ \InputMode{\Pair{M}{N}}
    \mid \Member{M}{P}, \Member{N}{Q}
    \}\\
  \VAL{\Disj{P}{Q}} &\equiv
    \{ \InputMode{\Inl{M}}
    \mid \Member{M}{P}
    \} \cup
    \{ \InputMode{\Inr{M}}
    \mid \Member{M}{Q}
    \}\\
  \EXP{P} &\equiv
    \{ \InputMode{M}
    \mid \Eval{M}{M'}, \VAL{P}(\InputMode{M'})
    \}
\end{align*}

The material interpretation of the hypothetical judgment is crucial in the
realizability model; this is because only the verification objects (i.e.
canonical witnesses) are given. All the non-canonical witnesses are explained
via computation, and the use of logical consequence instead of material
consequence in the explanation of $\VAL{\Imp{P}{Q}}$ would have been disastrous.

For instance, it should be the case that $\Lam{x}{\Pair{\It}{\It}}$ is a witness
of $\Imp{\False}{\True}$. To see if this is the case in the
model, let us translate this into a concrete statement:
\begin{gather}
  \Member{\Lam{x}{\Pair{\It}{\It}}}{\Imp{\False}{\True}}\\
  \Eval{\Lam{x}{\Pair{\It}{\It}}}{\Lam{x}{\Pair{\It}{\It}}}, \quad
    \VAL{\Imp{\False}{\True}}(\InputMode{\Lam{x}{\Pair{\It}{\It}}})\\
  \GenJ{x}{
    \HypJ{\Member{\Pair{\It}{\It}}{\True}}{\Member{x}{\False}}
  }\\
  \GenJ{x}{
    \HypJ{\Member{\Pair{\It}{\It}}{\True}}{\EXP{\False}(\InputMode{x})}
  }\\
  \GenJ{x}{
    \HypJ{\Member{\Pair{\It}{\It}}{\True}}{\Eval{x}{M}, \VAL{\False}(\InputMode{M})}
  }
\end{gather}

And at this time, we may discharge the entire hypothetical judgment, since we
know that the unary relation $\VAL{\False}$ is empty, and so there can be no
such $M$.

If we did not have the material consequence at our disposal, then this statement
would not have been valid, unless we were to eschew the verificationist meaning
explanation and also add ``use'' rules (i.e.\ direct eliminations) for each
proposition in addition to the verification rules.

\subsection{Type Theories as Binary Logical Relations}

The unary logical relations express exactly the content of a realizability
model, but they still do not yield a theory of sets which is sufficient for
reasoning about mathematical objects, which have extensional identity. In order
to consider the equality of sets and witnesses, the unary logical relations
are replaced with binary ones (sc.\ partial equivalence relations), as follows:

\begin{gather*}
  \infer={
    \EqSet{A}{B}
  }{
    \Eval{A}{A'} &
    \Eval{B}{B'} &
    \InputMode{\VAL{A'}}\equiv\InputMode{\VAL{B'}}
  }\qquad
  \infer={
    \IsSet{A}
  }{
   \EqSet{A}{A}
  }\\
  \infer={
    \EqMember{M}{N}{A}
  }{
    \Eval{A}{A'} &
    \Eval{M}{M'} &
    \Eval{N}{N'} &
    \VAL{A'}(\InputMode{M'}, \InputMode{N'})
  }\qquad
  \infer={
    \Member{M}{A}
  }{
    \EqMember{M}{M}{A}
  }
\end{gather*}

In order to make an important point about functionality, we will define
intuitionistic existential and universal quantification rather than their
special cases, conjunction and implication.
\begin{align*}
  \VAL{\True} &\equiv \{(\InputMode{\It}, \InputMode{\It})\}\\
  \VAL{\False} &\equiv \{\}\\
  \VAL{\Forall{A}{x}{B}} &\equiv
    \{ (\InputMode{\Lam{x}{E}}, \InputMode{\Lam{x}{E'}})
    \mid \GenJ{y,z}{
      \HypJ{\EqMember{[y/x]E}{[z/x]E'}{[y/x]B}}{\EqMember{y}{z}{A}}
    }
    \}\\
  \VAL{\Exists{A}{x}{B}} &\equiv
    \{ (\InputMode{\Pair{M}{N}}, \InputMode{\Pair{M'}{N'}})
    \mid \EqMember{M}{M'}{A}, \EqMember{N}{N'}{[M/x]B}
    \}\\
  \VAL{\Disj{A}{B}} &\equiv
    \{ (\InputMode{\Inl{M}}, \InputMode{\Inl{N}})
    \mid \EqMember{M}{N}{A}
    \} \cup
    \{ (\InputMode{\Inr{M}}, \InputMode{\Inr{N}})
    \mid \EqMember{M}{N}{B}
    \}\\
  \EXP{A} &\equiv
    \{ (\InputMode{M}, \InputMode{N})
    \mid \Eval{M}{M'}, \Eval{N}{N'}, \VAL{A}(\InputMode{M'}, \InputMode{N'})
    \}
\end{align*}

Now, in contrast to the treatment of the quantifiers in formal intuitionistic
logic (or in the Mitchell-B\'enabou language of a topos
\cite{maclane-moerdijk}), in this setting it is part of their meaning that their
verifications should respect the equality of the domain of discourse; this
constraint is called functionality, and reflects the fact that universal
quantification is reconstructed as a more general form of implication. Within
the theory of sets, there is simply not a quantifier which expresses
non-functional generality; in this way, contrary to the state of affairs in
\BISH{} \cite{bishop:1967}, the theorem of choice is in fact verified by a
choice \emph{function}, not merely a choice \emph{operation}.

With the extension to the binary logical relation, we now properly treat the
equivalence of propositions (sets) and of witnesses, and we also have a
definitive answer to the question, ``What is the purpose of adding a language of
types and witnesses to the existing system of judgments and their
demonstrations?''

The judgments and demonstrations are the activity of the subject in performing
and experiencing mathematics. On the other hand, the theory of sets that we have
defined is, to my mind, the correct level at which to \emph{do} mathematics,
where objects are concrete and have an extensional identity. By guaranteeing
pervasive functionality, we have embedded in the rich world of intuitionistic
mathematics a haven invulnerable to the paradoxes that arise from failures of
extensionality, such as Diaconescu's theorem \cite{martin-lof:2009, sterling:diaconescu}.

\section{Related Discussion}

In the logical framework which forms the basis of \emph{Practical Foundations
for Programming Languages} \cite{PFPL}, Robert Harper treats both logical
consequence $\JJ_1\vdash\JJ_2$ and material consequence $\JJ_1\vDash\JJ_2$,
which express derivability and admissiblity respectively.

The material consequence (and its open-ended interpretation as a mapping from
demonstrations of the antecedent to demonstrations of the consequent) formed the
backbone of Zeilberger, Harper and Licata's work on higher-order focused calculi, which
provide a convincing abstract notation for the traces of verification and use
acts in a logic which mixes the verificationist and pragmatist meaning
explanations \cite{zeilberger:thesis, licata-zeilberger-harper:focusing,
zeilberger:2008}.

\subsubsection*{Acknowledgements}

Thanks to Mark van Atten, Robert Harper, Clarissa Littler, and Danny Gratzer for
helpful conversations about hypothetical judgment, elimination rules and the bar
theorem.

\nocite{*}
\bibliographystyle{plain}
\bibliography{refs}

\end{document}